\documentclass[secnumarabic,nobibnotes,aps,prl,superscriptaddress,floatfix]{revtex4}

\usepackage{multirow}
\usepackage{graphicx}
\usepackage{amsmath}
\usepackage{enumerate}
\usepackage{epstopdf}
\usepackage{times}
\usepackage{color}
\usepackage{bm}
\usepackage{comment}

\begin{document}

\title{Phase Transition and Thermal Order-by-Disorder 
in the Pyrochlore Quantum Antiferromagnet Er$_2$Ti$_2$O$_7$:
A High-Temperature Series Expansion Study}

\author{J. Oitmaa }
\affiliation{School of Physics, The University of New South Wales,
Sydney 2052, Australia}
\author{R. R. P. Singh}
\affiliation{Department of Physics, University of California Davis, CA 95616, USA}
\author{A. G. R. Day}
\affiliation{Department of Physics and Astronomy, University of Waterloo, Waterloo, ON, N2L 3G1, Canada}
\author{B. V. Bagheri}
\affiliation{Department of Physics and Astronomy, University of Waterloo, Waterloo, ON, N2L 3G1, Canada}
\author{M. J. P. Gingras}
\affiliation{Department of Physics and Astronomy, University of Waterloo, Waterloo, ON, N2L 3G1, Canada}
\affiliation{Perimeter Institute for Theoretical Physics, 31 Caroline North, Waterloo, ON, N2L 2Y5, Canada}
\affiliation{Canadian Institute for Advanced Research, 180 Dundas Street West, Suite 1400, Toronto, ON, M5G 1Z8, Canada}

\date{\rm\today}

\begin{abstract}
Several rare earth magnetic pyrochlore materials are well
modeled by a spin-$1/2$  quantum Hamiltonian with anisotropic exchange parameters $J_s$.
 For the Er$_2$Ti$_2$O$_7$ material, the $J_s$ were recently determined from high-field inelastic
neutron scattering measurements. Here, we perform high-temperature ($T$)
series expansions to compute the thermodynamic properties
of this material using these $J_s$. 
Comparison with  experimental data show that the model describes the material very well including the
finite temperature phase transition to an ordered phase at $T_c\approx 1.2$ K. 
We show that high temperature
expansions give identical results for different ${\bm q}=0$ $xy$ 
order parameter susceptibilities up to $8^{\rm th}$ order in $\beta\equiv 1/T$ (presumably to all
orders in $\beta$). 
Conversely, a non-linear susceptibility related to the $6^{\rm th}$ power of the order parameter
reveals a thermal order-by-disorder selection of the {\it same} non-colinear ``$\psi_2$ state''
as found in Er$_2$Ti$_2$O$_7$.

\end{abstract}


\maketitle

{\it Order-by-disorder} (ObD) is a beautiful concept of central importance in the field of frustrated magnetism.~\cite{Villain,Shender}
Saddled with large accidental degeneracies, a subset of states, those that support the {\it largest}
quantum and/or thermal fluctuations, may be selected to form true long-range order,
thus turning on its head the conventional wisdom of ``less fluctuations lead to more order''.
While ObD has been discussed theoretically for over thirty years, 
and proposed to be at play in a number of experimental settings, 
most recently in cold atom systems,~\cite{ObD_cold_atoms} 
convincing demonstrations of ObD  in real materials have remained scarce.~\cite{Yildirim,Savary}
The main reason for the paucity of confirmed ObD material examplars is that the classical degeneracies
are typically not sufficiently symmetry-protected to rule out that 
some weak energetic perturbations
are responsible for stabilizing the observed long-range order.
An exception is the long-range order found in the $XY$ pyrochlore antiferromagnet Er$_2$Ti$_2$O$_7$.~\cite{Champion_PRB,Ruff_ETO_PRL,Poole}
Two groups \cite{Savary,Zhitomirsky} 
have recently put forward a strong case for a robust {\it highly protected} classical
degeneracy in that system, hence making the case for ObD much more compelling than in any previous examples.

\begin{figure}[ht]
\centering
\includegraphics[width=12cm,angle=0]{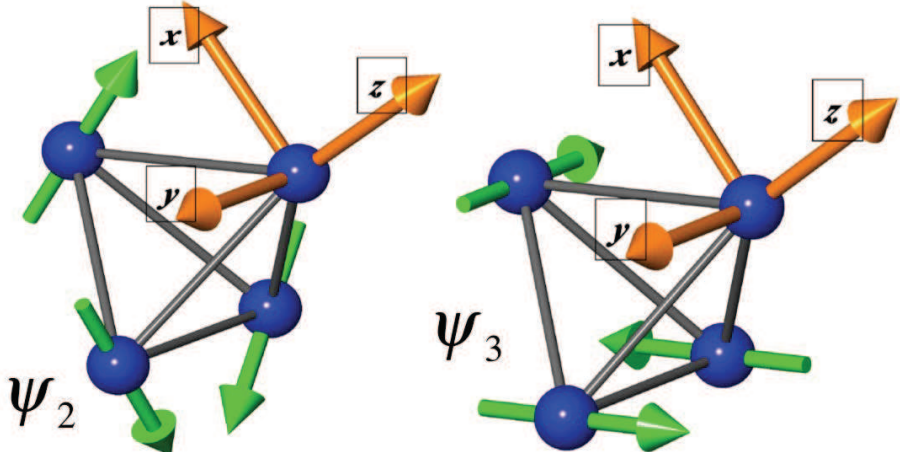}
\caption{The pyrochlore lattice can be described as a face-centered cubic lattice of elementary tetrahedra units.
In $\psi_2$ (left) and $\psi_3$ (right), each elementary tetrahedron has the shown spin configuration with 
moments along the local $x$  and $y$ axis for $\psi_2$ and $\psi_3$, respectively.
}
\label{fig:psi2andpsi3}
\end{figure}

In an $XY$ pyrochlore such as Er$_2$Ti$_2$O$_7$, the 
ordered moments lie on average in an $xy$  plane perpendicular to their local $[ 111 ]$ cubic direction.
At the classical level, energetics further require  a vanishing net magnetic moment on
each elementary tetrahedron of the pyrochlore lattice. 
Particularly interesting is the fact that among all classically degenerate ground states that 
satisfy such zero tetrahedral moment configuration, the material orders at a critical temperature $T_c \approx 1.2$ K
in a ground state 
that breaks a {\it discrete symmetry} within the local $[111]$ $xy$ plane -- the so-called $\psi_2$ basis state of the
 $\Gamma_5$ irreducible representation (see Fig. \ref{fig:psi2andpsi3}).~\cite{Champion_PRB,Poole}
Since  the original determination \cite{Champion_PRB} of the long-range order in Er$_2$Ti$_2$O$_7$,  
two rather puzzling questions had been identified:
Firstly, how could the system order into a state that does not minimize the  dipolar
interactions, the so-called Palmer-Chalker state \cite{Palmer_Chalker} which has each spin in its local $xy$ plane?
Secondly, what mechanism may lead to the $\psi_2$ selection, as opposed to the other
 $\psi_3$ basis vector of the two-dimensional $\Gamma_5$ manifold (see Fig.\ref{fig:psi2andpsi3}), 
or even an arbitrary superposition of $\psi_2$ and $\psi_3$ arising from a spontaneous $XY$ ($U(1)$) global symmetry breaking?

Earlier studies had shown thermal~\cite{Champion_PRB,Champion_JPCM,Stasiak}  and quantum~\cite{Stasiak} 
 ObD selecting a $\psi_2$ state in a simplified pyrochlore antiferromagnetic $XY$ model.
More recent work  found that anisotropic exchange can efficiently compete with dipolar interactions and lower the energy 
of  $\Gamma_5$ below that of the Palmer-Chalker state.~\cite{McClarty_ETO}
Building on these results, Savary {\it et al.} \cite{Savary} showed that the classical degeneracies within  
$\Gamma_5$ are in fact immune to anisotropic bilinear spin-spin interactions of arbitrary form and range, leaving
quantum ObD (q-ObD) as essentially \cite{but_VCFE} the only plausible mechanism explaining
the $\psi_2$ low-temperature state in Er$_2$Ti$_2$O$_7$.
A similar conclusion was reached in Ref.~[\onlinecite{Zhitomirsky}].
The possible occurence of ObD in Er$_2$Ti$_2$O$_7$ represents a potentially significant result in
the field of highly frustrated magnetism.

While the arguments of Ref.~\cite{Savary} as per the symmetry-protection
of degeneracy are compelling, there is still reason to worry whether the ordering mechanism
 in Er$_2$Ti$_2$O$_7$ has indeed been fully unveiled.
In particular, one may ask whether the model of Ref.~\cite{Savary}
 describes accurately the thermodynamic behavior of Er$_2$Ti$_2$O$_7$ close to the transition and predict a 
$T_c \sim 1.2$ K in accord with experiment.
For example,  since this was not investigated in Ref.~\cite{Savary},
a concern one might have is whether the model, for ignoring
the long-range part of the dipole-dipole interactions \cite{dipoles_in_nn} 
 and other perturbations, does display
a thermal ObD at $T_c$ in the correct $\psi_2$ state, rather than $\psi_3$, 
which would then be inconsistent with experiments.~\cite{double_transition}
Conversely, one may ask whether the experimentally observed $\psi_2$
state at low $T$ is the true ground state of the material or a metastable relic of the
thermal ObD at $T_c$.~\cite{supp_mater}
Finally,  the recent observation that there exists a tendency for
 rare-earth ions (e.g. R=Er$^{3+}$ ) in R$_2$Ti$_2$O$_7$  pyrochlore oxides 
to occupy the Ti$^{4+}$  site at the 1\% level,~\cite{Ross_stuff}
hence generating effective random magnetic interactions, 
also raises concerns whether a plausible q-ObD at low temperatures smoothly merges to its thermal variant at $T_c$.

The concerns above can only be alleviated by directly addressing, as we do in this paper, whether 
the model of  Ref.~[\onlinecite{Savary}]
describes well the thermodynamic behavior of Er$_2$Ti$_2$O$_7$ above $T_c$.
To do so, we use high-temperature expansions (HTE) and crystal-field theory to study the magnetic
specific heat and susceptibility of the model.
 We also calculate order-parameter susceptibilities for
$\psi_2$ and $\psi_3$, finding that the model displays a continuous phase transition at
a $T_c\approx 1.2 $ K close to the experimental value. 
By calculating a non-linear susceptibility, we show that $\psi_2$ order is indeed selected by thermal ObD.
These results imply that the long-range part of the
 dipolar interactions neglected in Ref.~[\onlinecite{Savary}] are not important above $T_c$  \cite{dipoles_in_nn}
and that the model of Ref.~[\onlinecite{Savary}] is quantitatively accurate.
We are thus rather confident that ObD, both thermal and quantum, 
cooperate in Er$_2$Ti$_2$O$_7$ to select $\psi_2$ over the whole  temperature range
 $0<T \le T_c$.

{\it Model \& method} $-$
In a number of pyrochlore oxides \cite{Gardner}, 
the single-ion crystal-field magnetic doublet
ground state is separated from the lowest excited crystal-field energy levels by an energy gap 
$\Delta$ that is large compared to the microscopic interactions, ${\cal H}_{\rm mic}$, between the ions. 
This is the case for Er$_2$Ti$_2$O$_7$ for which ${\cal H}_{\rm mic} \sim 1$ K while $\Delta \sim 75$ K.
In such a case, one can use an effective spin-$1/2$ Hamiltonian ${\cal H}$ with 
bilinear anisotropic couplings, $J_s$,  to describe the interactions between ions, and where
${\cal H}$ is the projection of ${\cal H}_{\rm mic}$ 
onto the Hilbert space spanned by the single-ion ground doublets.
On symmetry grounds, the nearest-neighbor ${\cal H}$ can be 
parametrized by four exchange parameters as follows:
\begin{eqnarray}
{\cal H}&=&\sum_{\langle i,j\rangle} \{ J_{zz}S_i^z S_j^z -J_\pm (S_i^+ S_j^- +S_i^-S_j^+)
         + J_{\pm\pm} [\gamma_{ij}S_i^+S_j^+ + \gamma_{ij}^* S_i^-S_j^-]
         + J_{z\pm} [S_i^z(\zeta_{ij}S_j^+ +\zeta_{ij}^*S_j^-) + i \leftrightarrow j] \}
\label{Hami}
\end{eqnarray}
Here, $\langle i,j\rangle$ refers to nearest-neighbor sites of the pyrochlore lattice,
$\gamma_{ij}$  is a $4 \times 4$ complex unimodular matrix, and $\zeta=-\gamma^*$ \cite{Savary,Ross}.
The $\hat z$ quantization axis is along the local $[111]$ direction,
and $\pm$ refers to the two orthogonal local directions. 
The $J_s$ were determined from fits to inelastic neutron scattering spectra in the field polarized state.~\cite{Savary,Js_values}
The magnetic properties of the system are described by the Zeeman Hamiltonian, 
${\cal H}_{\rm Z} = -{g_{\rm L} \mu_{\rm B}} \sum_{i}  {\bm J}_i \cdot {\bm B}  $
 added to ${\cal H}$, 
where ${\bm J}$ is the $J=15/2$ angular momentum operator of  Er$^{3+}$, ${\bm B}$ is the applied 
magnetic field,  $\mu_{\rm B}$ is the Bohr magneton and $g_{\rm L}=6/5$ is the Er$^{3+}$ Land\'e factor.~\cite{Hz_form}
In this paper we investigate the thermodynamic properties of ${\cal H}$ above and near $T_c$ using
HTE.~\cite{Series_book}

We have computed
 series for the following quantities:
(1) log of the partition function, $\ln{Z}$, from which heat capacity and entropy are readily calculated; 
(2) uniform susceptibility as a linear response to a static applied external magnetic field;
(3) linear order parameter susceptibilities, $\chi_{xx}$ and $\chi_{yy}$, corresponding to $\psi_2$ and $\psi_3$ order, respectively; 
and 
(4) non-linear (4$^{\rm th}$ and 6$^{\rm th}$ order)
order parameter cumulants associated with $\psi_2$ and $\psi_3$ long-range order. 
We discuss below the reason for calculating  non-linear susceptibilities from these cumulants.~\cite{supp_mater}

\begin{figure}[ht!]
\begin{center}
\includegraphics[width=12cm,angle=0]{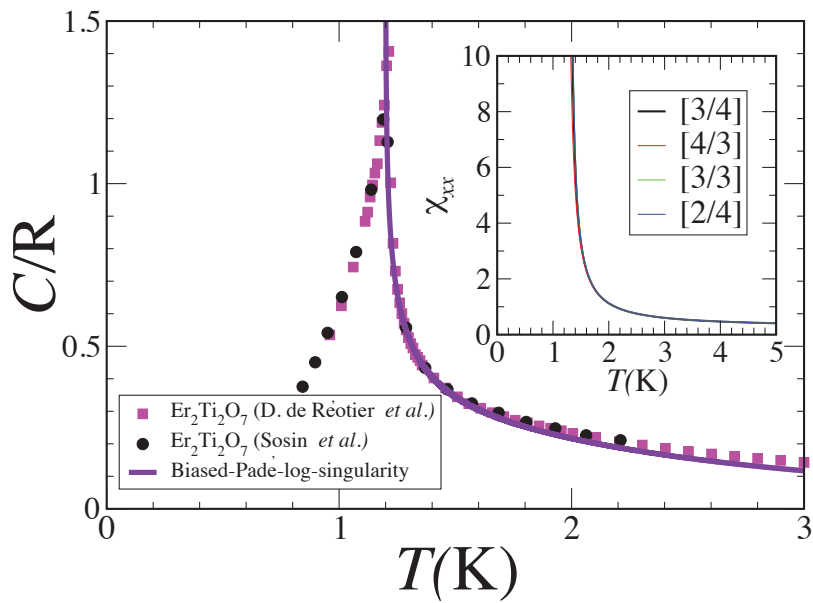}
\caption{\label{fig:spec_heat}
Thermodynamic behavior of model \eqref{Hami} compared to experimental data.
The main panel shows 
specific heat data for Er$_2$Ti$_2$O$_7$ compared with Pad\'e approximants for the high
temperature series expansions biased to have a logarithmic
singularity at $T_c$.
The experimental data are from Dalmas de R\'eotier {\it et al.} \cite{Dalmas_ETO} and
Sosin {\it et al.} \cite{Sosin}.
The inset shows plots of the order parameter susceptibility ($\chi_{xx}=\chi_{yy}$) calculated from high temperature expansions
for various dlog Pad\'e approximants (see text).
}
\end{center}
\end{figure}

{\it Demonstrating the validity of ${\cal H}$} $-$
We first show evidence that the Hamiltonian \eqref{Hami} is consistent with a phase transition to long-range order in
the $xy$ components of the spins at a critical temperature $T_c \sim 1.2$ K.
To do so, we calculate high temperature series expansions for the $xy$ order-parameter susceptibilities.
We apply a field along the local $x$ ($y$)-axis at all sites, and calculate the linear
response order-parameter susceptibility, $\chi_{xx}$ ($\chi_{yy}$). 
Detailed expressions can be found in the supplementary material ~\cite{supp_mater}.
As discussed in Refs. \cite{Savary,Zhitomirsky},
there is an emergent continuous symmetry in the model \eqref{Hami}, 
which is only weakly lifted by higher order effects beyond
classical ground state energetics. 
In the notation of Ref. \cite{Savary}, this classical degeneracy can be parameterized by a continuous angle  $\alpha$. 
Then, $\chi_{xx}$ is the susceptibility for $\alpha=0$ order (i.e. $\psi_2$ order)
while $\chi_{yy}$ is the susceptibility for $\alpha=\pi/6$ order (i.e. $\psi_3$ order). 
We find that the two linear susceptibilities have {\it identical} high temperature series expansions to
the order calculated (8$^{\rm th}$ order in $\beta$). We believe that this result is true to all orders and
the selection of order within the $\Gamma_5$  manifold must therefore only be manifest in non-linear 
order parameter susceptibilities.
We return to this matter later.
We study the singularities of the order-parameter susceptibility using d-log Pad\'e approximants.~\cite{Series_book}
Various  estimated $T_c$  and critical exponents $\gamma$ from near diagonal d-log Pad\'e values,
expressed in the form of $(L,M;T_c,\gamma$) sets,  are
$(3,4;1.26,1.21)$, 
$(4,3;1.25,1.71)$,
$(3,3;1.33,1.19)$ and $(2,4;1.28,1.14)$ where $T_c$ is in Kelvin.
Although the convergence is not excellent for this short series, we  nevertheless conclude that the
transition temperature is $T_c=1.2\pm 0.1$ K. Given the large uncertainty in $T_c$ we cannot
make a reliable estimate for $\gamma$, but its $\gamma = 1.3 \pm 0.4$ value 
is consistent with $\gamma$ values known for $3$-dimensional spin models.
A plot of the order parameter susceptibility ($\chi_{xx}=\chi_{yy}$) versus  
temperature is shown in the inset of Fig. \ref{fig:spec_heat}.

Next, we turn to a calculation of the specific heat, $C(T)$, and its comparison with experiments.
We have calculated specific heat using both Numerical Linked Cluster (NLC) method \cite{Rigol,Applegate,Hayre} and
HTE. The two methods agree well at $T>2$ K. At lower
temperatures, the HTE method is better as it allows us to analyze the behavior near
the phase transition, where the correlation length diverges.
Since we expect the system to display a three-dimensional $XY$ universality class,
for which the specific heat exponent $\alpha$ is known to be very close to zero,
we bias the analysis of the specific heat series to have a log singularity at $T_c=1.2$ K.
The biased analysis shows good convergence with several approximants found to be very close to each other. 
A representative
plot is shown together with experimental data in Fig. \ref{fig:spec_heat}, where an 
excellent agreement with  experimental data is seen.
The phonon contributions are clearly seen to be 
negligible below $2$ K. 
The magnetic entropy  above $T_c$ can be found in the supplementary material.~\cite{supp_mater}
 Its value at $T_c$ is reduced from the infinite temperature value by less than $50$ percent,
not atypical of $3$-dimensional critical points.

We now turn to the uniform magnetic susceptibility,
which has been measured up to $300$ K.
As shown in Fig.~4, this data shows a crossover from a high temperature Curie constant
of $11.5 \text{ emu/mol}\cdot \text{K}$
to a low temperature Curie constant of $2.48 \text{ emu/mol}\cdot \text{K}$, reflecting
the evolution of the material from a $J=15/2$ system to an effective spin-$1/2$ system. To understand
this susceptibility data, we use HTE to calculate the susceptibility for the effective spin-$1/2$ model
with  $g_\perp$ and $g_{zz}$ $g$-tensor components \cite{Hz_form}  from Ref.~[\onlinecite{Savary}],
but also the full single-ion susceptibility, $\chi_{\rm s.i.}$, 
obtained by including all the crystal-field states of the crystal-field Hamiltonian
of Er$_2$Ti$_2$O$_7$.~[\onlinecite{Bertin}]
The latter is obtained by 
treating the infinitesimal magnetic field ${\bm B}$ that 
couples to the non-interacting rare-earth 
ion with second order (degenerate) perturbation theory.~\cite{White_book}
 The single-ion susceptibility ($\chi_{\rm s.i.}$) then takes the form:
\begin{eqnarray}
\chi_{\rm s.i.}&=& {g_{\rm L}}^2 \mu_0 \mu_B^2 N_{\rm A} 
\frac{\sum_n  [\beta (E_n^{(1)})^2-2E_n^{(2)}] e^{-\beta E_n^{(0)}}}
{\sum_n e^{-\beta E_n^{(0)}}}.
\label{chivv}
\end{eqnarray}
Here $E_n^{(0)}$ is the energy of the $n$-th state of 
$H_{\mathrm{CF}}$ while $E_n^{(1)}$ and $E_n^{(2)}$ are  the first 
and second order perturbation theory coefficient of the energy of 
the $n^{\rm th}$ state, respectively.~\cite{White_book,Jensen_book}
 $N_{\rm A}$  is the Avogadro number and $\mu_0$ the vacuum permeability. 

\begin{figure}[ht]
\begin{center}
\includegraphics[width=10cm]{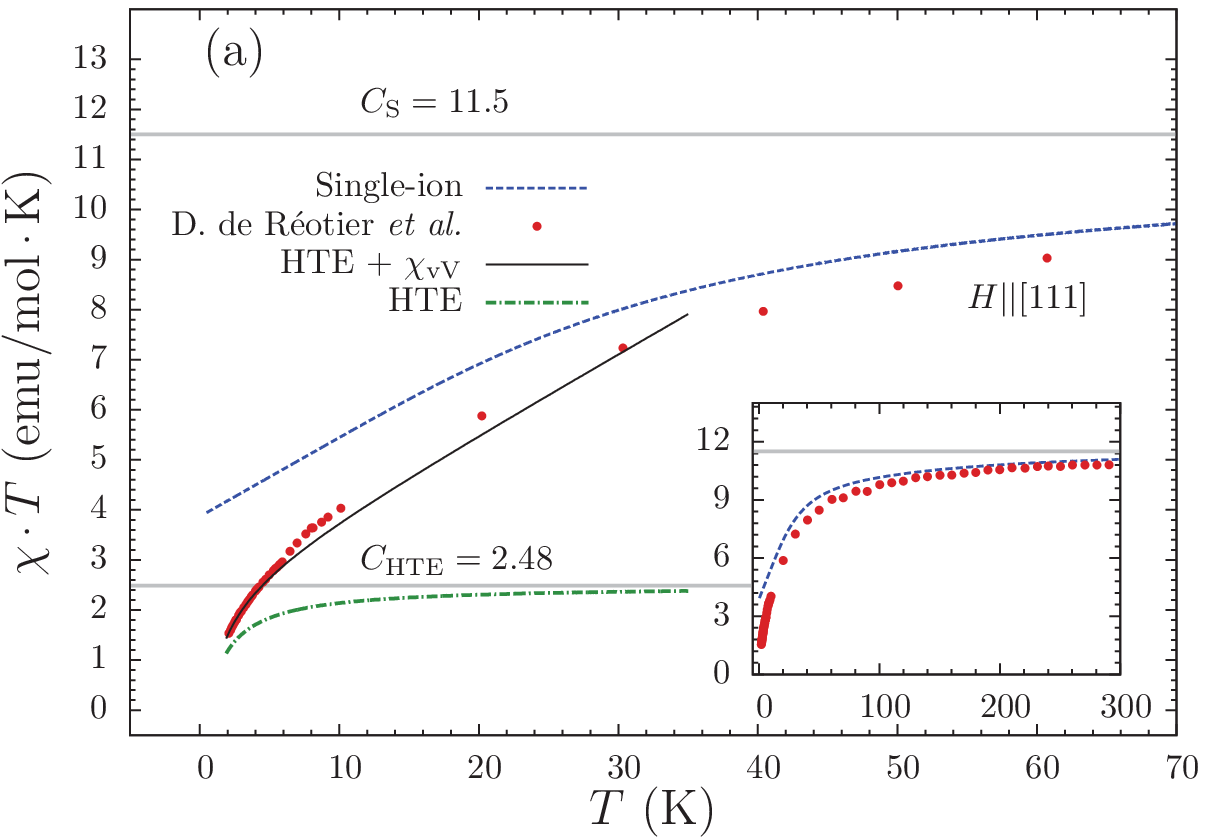}
\includegraphics[width=10cm]{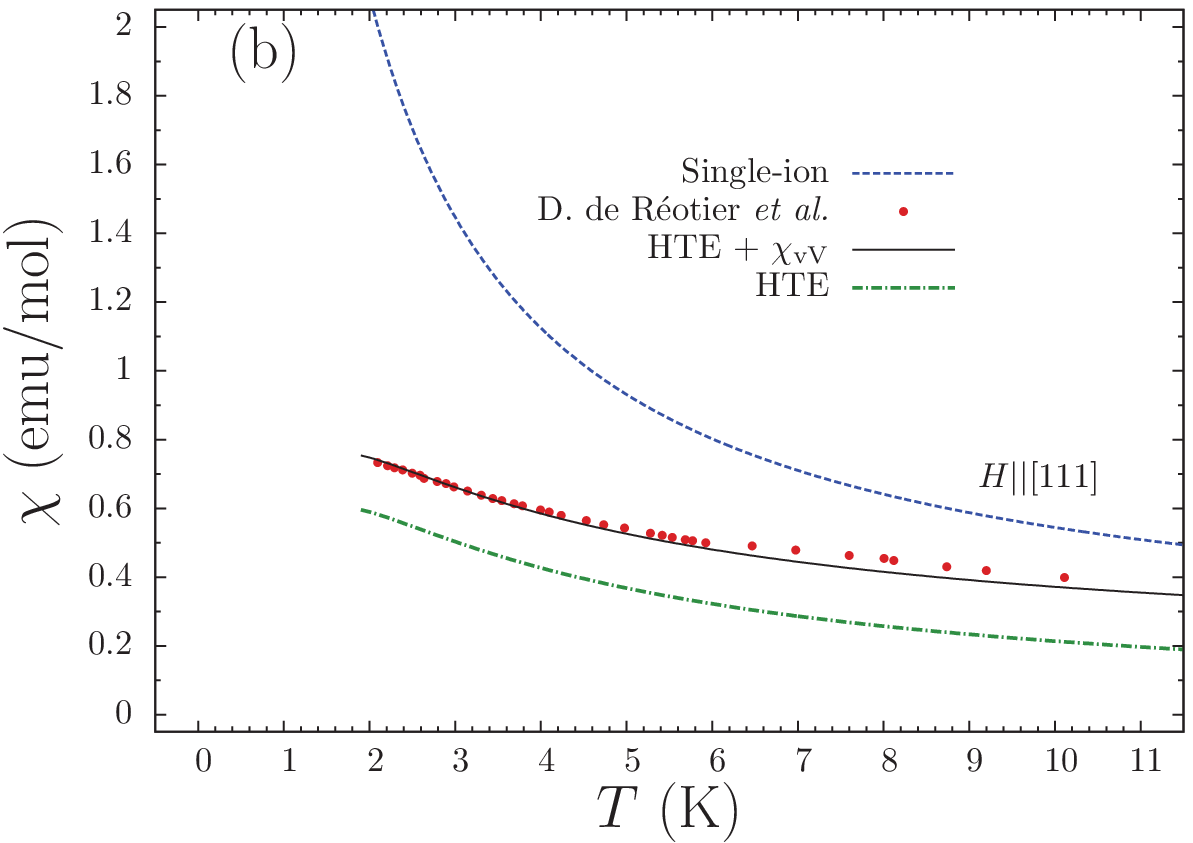}
\caption{(a) $[111]$ zero field uniform susceptibility, $\chi$,  times temperature, $T$,  versus $T$.
The dashed blue line corresponds to the single-ion calculations which agree with
the experimental data of Dalmas de R\'eotier {\it et al.} \cite{Dalmas_ETO}
at high temperatures   (see inset).
The top horizontal gray line corresponds to the calculated Curie constant,
$C_{\text{S}}=11.5 \text{ emu/mol}\cdot \text{K}$.
The dashed green line corresponds to the 6$^{\rm th}$ order HTE susceptibility using the Hamiltonian \eqref{Hami}.
Its corresponding Curie constant is also shown as a gray line
with value of $C_{\text{HTE}}=2.48 \text{ emu/mol}\cdot \text{K}$.
The van Vleck susceptibility, $\chi_{\mathrm{vV}}=0.158$  emu/mol, originating
from the admixing of the ground doublet with excited crystal-field states
and obtained from the single-ion calculation, is added to the HTE susceptibility to yield the black curve.
(b) The low temperature ($T \lesssim 10$ K)
susceptibility is reasonably well accounted by the HTE susceptibility corrected with the van Vleck susceptibility.
}
\label{fig:suscept}
\end{center}
\end{figure}

The inset of Fig.~3(a) shows a comparison of the measured susceptibility
with $\chi_{\rm s.i.}$.
The agreement at high temperatures is apparent. 
The increasing disagreement 
between the experimental $\chi$ and the calculated $\chi_{\rm s.i.}$ as
$T$ is reduced
is caused by the progressive development of {\it antiferromagnetic} correlations
which decrease the uniform susceptibility.
The main plot shows that below about $70$ K, 
$\chi_{\rm s.i.}$ 
 deviates significantly from the measured values. 
Also shown in Fig.~3(a) and Fig.~3(b) are comparisons 
with the susceptibility calculated by HTE for
the projected spin-$1/2$ model \eqref{Hami} taken alone.
 The latter saturates to a Curie law $C_{\rm HTE}/T$ form with
$C_{\text{HTE}}=2.48 \text{ emu/mol}\cdot \text{K}$ at high temperatures. 
The HTE expansion considers AF correlations within the low-energy Hilbert space spanned
by the single-ion crystal-field doublets of interacting Er$^{3+}$ ions.
These HTE calculations ignore the residual $T\lesssim 10$ K {\it temperature-independent}
contribution coming from the excited crystal-field states, which
is the so-called van Vleck susceptibility, $\chi_{\rm vV}$.~\cite{White_book,Jensen_book}
To compare experimental data with model calculations,
we thus need to add  $\chi_{\rm vV}$ to the HTE calculation results.
As seen in Fig.~4(b), one obtains a very good agreement with the experimental data
once $\chi_{\rm vV}$ is added.

{\it Thermal ObD  into $\psi_2$}  $-$ 
Having demonstrated
that the model in Eq.~\eqref{Hami} describes the thermodynamic
properties of the material above $T_c$ accurately, we now turn to the
central question of thermal ObD at $T_c$. We have already shown that the linear
order-parameter susceptibility fails to distinguish 
between $\psi_2$ and $\psi_3$ order upon approaching $T_c$ from 
the paramagnetic phase.
To further address this issue, we need to calculate non-linear
order-parameter susceptibilities constructed from
equal-time $4^{\rm th}$ and $6^{\rm th}$ power cumulants of the $\psi_2$ and $\psi_3$ order parameters
in HTE. We find that the $4^{\rm th}$ power 
cumulant is also identical for the two cases. However, the
$6^{\rm th}$ power of the order parameter does distinguish between 
$\psi_2$ and $\psi_3$ order 
(see Table VI in the supplementary materials \cite{supp_mater}). 
The necessity to compute a non-linear susceptibility  that is $6^{\rm th}$ order in the order-parameter 
 to reveal the selection of $\psi_2$ vs $\psi_3$ can be
understood on the basis that a $6^{\rm th}$ order effective ``potential'', $V(\alpha)$, 
($V(\alpha) \sim g_6 \cos(6\alpha)$ where 
$g_6<0$ and $g_6>0$ selects $\psi_2$ and $\psi_3$, respectively)
in the Ginzburg-Landau free-energy
is dynamically generated by thermal and quantum fluctuations.~\cite{Savary,McClarty_ETO}
Starting at order $\beta^4$, 
the $6^{\rm th}$ non-linear  order susceptibility for $\psi_2$ order 
(or $\alpha=0$) gets larger than for $\psi_3$ for every $\beta^n$ ($n\ge 4$) order considered.~\cite{supp_mater}
This $6^{\rm th}$ order susceptibility exposes 
a thermal order by disorder that coincides with the pattern of order selected at $T=0$ in
the works of Savary {\it et al.}\cite{Savary} and  Zhitomirsky {\it et al.}\cite{Zhitomirsky}
(whose model is slightly different from the one we study).

{\it Conclusion} $-$ We have shown that the nearest-neighbor
spin Hamiltonian \eqref{Hami},
with exchange parameters determined from inelastic neutron
scattering,~\cite{Savary} 
describes the thermodynamic properties of Er$_2$Ti$_2$O$_7$ at $T>T_c$, 
including the continuous phase transition
at $T_c$ rather adequately. 
This was demonstrated through detailed comparison of  calculated and experimental 
specific heat ($C(T)$) and uniform susceptibility ($\chi(T)$) data.
While the nearest-neighbor part of the dipolar interactions 
is implicitly incorporated in the model,~\cite{dipoles_in_nn} it lacks such
terms beyond nearest neighbors. In spite of that, the thermodynamic
properties  are described quite well. Similar conclusions were
recently reached about a nearest-neighbor model determined
by inelastic neutron scattering for the material Yb$_2$Ti$_2$O$_7$.~\cite{Ross,Applegate,Hayre}

We have also shown that in the paramagnetic phase, 
the  linear order-parameter 
susceptibility does not reveal the lifting of the $\Gamma_5$ manifold degeneracy.
The selection of $\psi_2$ order is only evinced through a non-linear susceptibility 
that is $6^{\rm th}$ order in the order-parameter.
The thermal order-by-disorder identified in this paper 
was found to occur in the $\psi_2$ state,
in agreement with experiments \cite{Champion_PRB,Ruff_ETO_PRL,Poole} 
and with the quantum order-by-disorder  calculations of Refs.~[\onlinecite{Savary,Zhitomirsky}]. 
We thus conclude that the material Er$_2$Ti$_2$O$_7$ presents a unique and convincing 
case of {\it co-operating} quantum and thermal order-by-disorder in a frustrated quantum antiferromagnet.

\begin{acknowledgements}

We thank Behnam Javanparast, Anson Wong and Zhihao Hao for useful discussions.
This work is supported in part by NSF grant number  DMR-1004231 (RRPS),
the NSERC of Canada, 
the Canada Research Chair program (M.G., Tier 1) and by the Perimeter Institute for Theoretical Physics. 

\end{acknowledgements}

\newpage

\section{Supplemental Material}

This section provides supplemental material to the main part of our paper.
Firstly, we discuss the hypothetical situation when the thermal order-by-disorder can lead to a different long-range ordered state
than the quantum order-by-disorder.
Secondly, we provide some details regarding the high-temperature series expansion, in particular in regards to the
order parameter linear and non-linear susceptibilities. 
Then we show the results for the temperature dependence of the magnetic entropy,
$S(T)$, corresponding to the magnetic specific heat $C(T)=TdS/dT$ presented in the body of the paper.
Finally, 
we provide tables
for the high-temperature series expansion of various quantities of interest and referred to in the main text.

\subsection{Metastability of States Selected by Thermal Order-by-Disorder}
 
One may ask whether the long-range ordered state that is selected at the critical temperature
differs from the zero-temperature ground state selected by quantum fluctuations.
In case, thermal and quantum ObD are different,
there is a possibility of metastability of the thermal ObD state at
low temperatures.
Thus, the concern here has to do with the kinetics in the real material rather than thermodynamics.
Unlike natural proteins, extensively degenerate systems may lack funneled free-energy landscapes.
As a result, configuration-space pathways and dynamical access to states may play a
role in the specific long-range ordered  phases experimentally realized.
Ref.~[\onlinecite{Savary}] finds a zero-point energy stabilization
($\delta E_0$) of $\psi_2$ compared to $\psi_3$ of merely $\delta E_0 \sim 0.3$ $\mu$eV
 (a factor 1/30 of the reported smallest anisotropic exchange.~\cite{Savary})
One may then ask whether the neglected interactions {\it beyond} nearest neighbors,
in particular the dipolar interactions of magnitude 10 $\mu$eV $\sim 30 \delta E_0$,~\cite{dipoles_in_nn}
 may have instead led to a $\psi_3$ ground state via quantum ObD.
 Meanwhile,   {\it thermal} ObD, not considered in Ref.~[\onlinecite{Savary}],
may in fact be responsible for a transition into $\psi_2$ at $T_c$, both in the material and
in an amended version of the model of Ref.~[\onlinecite{Savary}]
that would incorporate interactions neglected therein.~\cite{double_transition}
Following such a thermal ObD  into $\psi_2$ in the material, a lower temperature
ordering into  $\psi_3$, or even in the Palmer-Chalker state, may be dynamically inhibited,
similar to what is found in Monte Carlo simulations of a pyrochlore $XY$ antiferromagnet with weak
dipolar interactions.~\cite{Stasiak,move_critical_dipole,Pinettes}
In such a scenario, the model of Ref.~\cite{Savary}
 would seemingly predict the  correct low-temperature state of Er$_2$Ti$_2$O$_7$  $-$ but for the wrong reasons.
To rule out the possibility of metastability of $\psi_2$ at low temperatures,
one could cool down to a high-field phase and gradually lower the field at low temperatures to assess whether
$\psi_2$ is still realized. To our knowledge, such an experiment has not been carried out.~\cite{thx_field}

\subsection {Notes on High Temperature Series Expansion for Er$_2$Ti$_2$O$_7$}
\label{supp:sec-HTE}

We consider the Hamiltonian
\begin{equation}
{\cal H} = {\cal H}_{\rm e}+H_{\rm Z} + H_x + H_y,
\end{equation}
Here ${\cal H}_{\rm e}$ is the exchange Hamiltonian defined in terms of
local spin variables and exchange constants
$J_{zz}$, $J_{z\pm}$, $J_{\pm}$ and $J_{\pm\pm}$. The next term
$H_{\rm Z}$ is the Zeeman term arising from the applied 
external field ${\bm B}$ which,
for a field along $[111]$, can be written in terms of local spin variables as
\begin{eqnarray}
{\cal H}_{\rm Z} & = & -B\ [ g_{zz} S_0^z -{1\over 3} g_{zz} (S_1^z+S_2^z+S_3^z) \\ \nonumber
    & & -{2\sqrt{2}\over 3}g_{xy}S_1^x	
   +  {\sqrt{2}\over 3} g_{xy}(S_2^x+S_3^x)
-\sqrt{{2\over 3}} g_{xy}(S_2^y-S_3^y) ]	.
\end{eqnarray}
Here the subscripts $0$, $1$, $2$ and $3$ denote different sublattices and one needs to sum over all sites.
The last two terms in the Hamiltonian are auxiliary field terms introduced for computational purposes.
\begin{equation}
H_x=-h_x \sum_i S_i^x
\end{equation}
and
\begin{equation}
H_y=-h_y \sum_i S_i^y.
\end{equation}
All spins are defined in their local basis.

High temperature series expansions are calculated for the 
logarithm of the zero-field
partition function $\ln{(Z_0)}$, with $Z_0$ given by
\begin{equation}
Z_0={\rm Tr} \exp{(-\beta {\cal H}_{\rm e})}.
\end{equation}
From $\ln{(Z_0)}$ the entropy and specific heat series
follow by simple differentiation.

The zero field uniform susceptibility is calculated as the second derivative of the
free energy with respect to the applied external magnetic field, $B$, 
with $h_x=h_y=0$.
\begin{equation}
\chi ={-1\over N\beta} {\partial^2\over \partial B^2} \ln{(Z(B))}\big|_{B=0} .
\end{equation}
In calculating $\chi$ from this equation, one obtains only the contribution from 
the interacting ions assumed to be in their single-ion crystal field ground states
and neglecting any contribution from excited crystal field levels.
To reach quantitative agreement with experiments, one needs to incorporate the 
residual low-temperature contribution to the susceptibility coming
from the excited crystal field states. This is the so-called van Vleck susceptibility.~\cite{White_book,Jensen_book}

To calculate the order-parameter susceptibility for $\psi_2$, we
set $B=h_y=0$ and calculate the second derivative of the free energy
with respect to $h_x$.
\begin{equation}
\chi_{xx} ={-1\over N\beta} {\partial^2\over 
\partial h_x^2} \ln{(Z(h_x))}\big|_{h_x=0}
\end{equation}
Similarly, for $\psi_3$, order we set $B=h_x=0$ and take
second derivative of free energy with respect to $h_y$
\begin{equation}
\chi_{yy} ={-1\over N\beta} {\partial^2\over 
\partial h_y^2} \ln{(Z(h_y))}\big|_{h_y=0}
\end{equation}
Evidently the series for $\chi_{xx}$ and $\chi_{yy}$ are identical.

Non-linear susceptibilities do not appear to have linked-cluster 
property. Instead, higher order zero field cumulants were calculated
by Linked cluster expansion. These are analogous to equal-time
structure factors (rather than zero-frequency susceptibilities).
Setting $B=h_x=h_y=0$, we define for $\alpha=x$ or $\alpha=y$
the order-parameter operator
\begin{equation}
M_\alpha \equiv \sum_i S_i^\alpha.
\end{equation}
Then, the cumulants, $C_{n,\alpha}$,  are:
\begin{equation}
C_{2,\alpha} \equiv \langle M_\alpha^2 \rangle,
\end{equation}
\begin{equation}
C_{4,\alpha} \equiv  \langle  M_\alpha^4 \rangle -3 \langle M_\alpha^2 \rangle^2,
\end{equation}
and
\begin{equation}
C_{6,\alpha} \equiv  \langle M_\alpha^6 \rangle - 15\langle M_\alpha^4\rangle  \langle M_\alpha^2\rangle +30 \langle M_\alpha^2\rangle^3. 
\end{equation}
Here, one should note that the bare moments $\langle M_\alpha^4 \rangle$ and $\langle M_\alpha^6 \rangle$
do not satisfy linked cluster property but the $C_{n,\alpha}$ cumulants do. 
Since $ \langle M_\alpha^2 \rangle$ and $\langle M_\alpha^4 \rangle$ are identical term by term, the
difference 
$\langle M_x^6 \rangle   - \langle M_y^6 \rangle $ is equal to the difference in the cumulants,
$ \langle C_{6,x}\rangle - \langle C_{6,y}\rangle$.

\newpage

\section{Temperature Dependence of the Magnetic Entropy, $S(T)$}
\label{supp:sec-entropy}

\begin{figure}
\begin{center}
\includegraphics[width=10cm,angle=270]{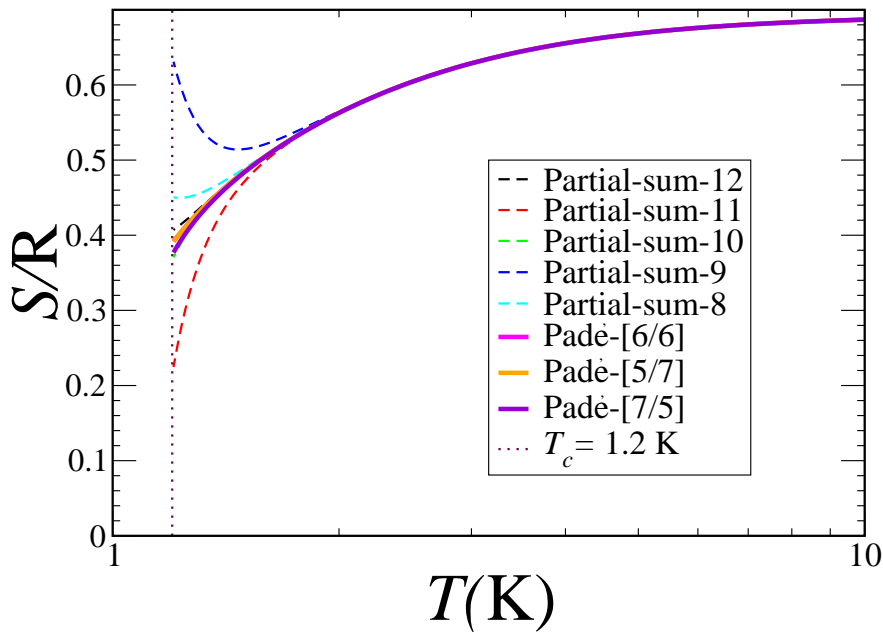}
\caption{\label{fig:entropy} 
A plot of the entropy function for the model. The dotted line shows the estimated critical point. The calculations
presented here are only informative above $T_c$.}
\end{center}
\end{figure}

The magnetic entropy function, $S(T)$, is shown in Fig. \ref{fig:entropy}. 
We show the partial sums of the series from order 8 to order 12 as well
as a few Pad\'e approximants. 
At the critical temperature the entropy per mole in units of the perfect gas constant $R$ is around $0.4$.
In other words, it is reduced from the infinite temperature value, $S_\infty=R\ln(2)$, by less than fifty percent. 
This is not unusual compared to typical three-dimensional phase transitions to long-range order.
This indicates that the system is not highly frustrated with the development of a strongly correlated regime
above $T_c$. A similar conclusion was reached in Ref.~[\onlinecite{Savary}] on the basis of a comparison between
the mean-field critical temperature of the model, $T_c^{\rm mf} \sim 2.3$ K, and the experimentally observed $T_c=1.2$ K

\subsection{High-Temperature Series for Various Quantities}
\label{supp:sec-Tables}

In this section, we provide the reader with tables of coefficients
forthe high-temperature series of quantities of interest.
Unless stated otherwise, the high-temperature series of a quantitity,
$Q(T)$, as a function of inverse temperature $\beta \equiv 1/T$ (with temperature $T$ in
Kelvin), is expressed in the form
\begin{eqnarray}
Q(T)   & = & \sum_{n=0}^{n_{\rm max}} a_n \beta^n.
\end{eqnarray}

\begin{table}
\caption{$\ln(Z)$ (per spin) series with only the two largest $J_s$ (in Kelvin); 
$J_\pm = 0.754$, $J_{\pm\pm} = 0.487$, $J_{zz}= 0$ and $J_{z\pm}=0$.
The label $E$ $\pm m$ means $10^{\pm m}$.  }
\centering 
\begin{tabular}{c c } 
\hline\hline 
Order $n$ & expansion coefficient $a_n$ \\ [0.5ex] 
\hline 
 0	&	$+0.693147181$ $E$ $+00$ 		 \\
 1	&	$+0.000000000$ $E$ $+00$ 		 \\
 2 	&	$+0.604263750$ $E$ $+00$  	  \\
 3 	&   	$+0.802114625$ $E$ $-01$			\\
 4	&	$-0.218372147$ $E$ $+00$ 			\\
 5	&	$-0.863347964$ $E$ $-01$ 		 \\
 6 	&	$+0.158902388$ $E$ $+00$  			\\
 7 	&	$+0.120287973$ $E$ $+00$			\\
 8	&     $-0.958592054$ $E$ $-01$ 		\\
 9 	&	$-0.118324998$ $E$ $+00$  	\\
 10 	&	$+0.361281401$ $E$ $-01$  		\\
 11 	&	$+0.728830750$ $E$ $-01$			\\
 12 	&	$+0.33501756$3 $E$ $-01$			\\ [1ex] 
\hline 
\end{tabular}
\label{table:free_energy_4} 
\end{table}

\begin{table}
\caption{$\ln(Z)$ (per spin) series with  all four $J_s$ (in Kelvin); 
$J_{\pm} = 0.754$, $J_{\pm\pm} = 0.487$, $J_{zz}= -0.290$ and $J_{z\pm}=-0.102$.
The label $E$ $\pm m$ means $10^{\pm m}$.  }
\centering 
\begin{tabular}{c c } 
\hline\hline 
Order $n$ & expansion coefficient $a_n$ \\ [0.5ex] 
\hline 
0	&	  $+0.693147181$ $E$  $+00$  \\
1	&	  $+0.000000000$ $E$  $+00$  \\
2	&  $+0.619951125$ $E$  $+00$  \\
3	&  $+0.604908776$ $E$ $-01$	\\
4	& $-0.257221456$ $E$  $+00$ 	\\
5	& $-0.662731421$ $E$ $-01$  	\\
6	&  $+0.223356105$ $E$  $+00$  	\\
7	&  $+0.103109973$ $E$  $+00$	\\
8	& $-0.200436766$ $E$  $+00$ 	\\
9	& $-0.120572677$ $E$  $+00$  	\\
10	&  $+0.186373242$ $E$  $+00$  	\\
11	&  $+0.113320221$ $E$  $+00$	\\
12	& $-0.155050053$ $E$  $+00$ 	\\ [1ex] 
\hline 
\end{tabular}
\label{table:free_energy_all} 
\end{table}

\begin{table}
\caption{Order parameter susceptibility $\chi_{xx}=\chi_{yy}$ per spin
with  all four $J_s$ (in Kelvin); 
$J_\pm = 0.754$, $J_{\pm\pm} = 0.487$, $J_{zz}= -0.290$ and $J_{z\pm}=-0.102$.
The label $E$ $\pm m$ means $10^{\pm m}$.  }
\centering 
\begin{tabular}{c c } 
\hline\hline 
 Order $n$ & expansion coefficient $a_n$ \\ [0.5ex] 
\hline 
0	&	$+0.250000000$ $E$  $+00$  	\\
1  	&	$+0.565500000$ $E$  $+00$  	\\
2 	&	$+0.856772872$ $E$  $+00$  	\\
3  	&	$+0.105521359$ $E$  $+01$	\\
4  	&	$+0.130532942$ $E$  $+01$  	\\
5  	&	$+0.181908741$ $E$  $+01$  	\\
6 	&	$+0.253686031$ $E$  $+01$  	\\
7  	&	$+0.320904312$ $E$  $+01$	\\
8  	&	$+0.386983904$ $E$  $+01$	\\ [1ex] 
\hline 
\end{tabular}
\label{table:free_energy_all} 
\end{table}

\begin{table}
\caption{Second order order-parameter $C_{2,x}$ cumulant   per spin
with  all four $J_s$ (in Kelvin); 
$J_\pm = 0.754$, $J_{\pm\pm} = 0.487$, $J_{zz}= -0.290$ and $J_{z\pm}=-0.102$.
Each coefficient is the same for the second order $C_{2,y}$ cumulant to the last digit shown.
The label $E$ $\pm m$ means $10^{\pm m}$.  }
\centering 
\begin{tabular}{c c } 
\hline\hline 
 Order $n$ & expansion coefficient $a_n$ \\ [0.5ex] 
\hline 
 0	& $+0.250000000000$ $E$  $+00$  	\\
 1	&  $+0.565500000000$ $E$  $+00$  	\\
 2	& $+0.936030375000$ $E$  $+00$  	\\
 3	& $+0.107532105831$ $E$ $+01$		\\
 4	& $+0.123006642254$ $E$ $+01$  		\\
 5	& $+0.179517883700$ $E$ $+01$  		\\
 6	& $+0.263550433751$ $E$ $+01$  		\\
 7	& $+0.325196289151$ $E$ $+01$		\\
 8	& $+0.374996212319$ $E$ $+01$	\\ [1ex] 
\hline 
\end{tabular}
\label{table:free_energy_all} 
\end{table}

\begin{table}
\caption{Fourth order order-parameter $C_{4,x}$ cumulant   per spin
with  all four $J_s$ (in Kelvin); 
$J_\pm = 0.754$, $J_{\pm\pm} = 0.487$, $J_{zz}= -0.290$ and $J_{z\pm}=-0.102$.
Each coefficient is the same for fourth order $C_{4,y}$ cumulant to the last digit shown.
The label $E$ $\pm m$ means $10^{\pm m}$.  }
\centering 
\begin{tabular}{c c } 
\hline\hline 
 Order $n$ & expansion coefficient $a_n$ \\ [0.5ex] 
\hline 
 0	&  $-0.125000000000$ $E$  $+00$ 	\\
  1	& $-0.113100000000$ $E$ $+01$ 		\\
  2	& $-0.532304971875$ $E$ $+01$ 		\\
  3	& $-0.172896459338$ $E$ $+02$		\\
  4	& $-0.437448147843$ $E$ $+02$ 		\\
 5	&  $-0.961820514195$ $E$ $+02$ 		\\
  6	& $-0.196495861437$ $E$ $+03$ 		\\
  7	& $-0.381391503061$ $E$ $+03$		\\
  8	& $-0.704597614446$ $E$ $+03$	\\ [1ex] 
\hline 
\end{tabular}
\label{table:free_energy_all} 
\end{table}

\begin{widetext}
\begin{table}
\caption{Sixth order order-parameter $C_{6,x}$ cumulant and $C_{6,y}$ cumulant   per spin
with  all four $J_s$ (in Kelvin); 
$J_\pm = 0.754$, $J_{\pm\pm} = 0.487$, $J_{zz}= -0.290$ and $J_{z\pm}=-0.102$.
The label $E$ $\pm m$ means $10^{\pm m}$. 
The last column shows the difference between the two cumulants.
Note the increasing difference with increasing $\beta^n$ order. }
\centering 
\begin{tabular}{c c c c} 
\hline\hline 
 Order $n$ & expansion coefficient $a_n$ ($C_{6,x}$-cumulant) & expansion coefficient $a_n$ ($C_{6,y}$-cumulant) 
& difference \\ [0.5ex] 
\hline 
 0	& $+0.250000000000$ $E$ $+00$  &	$+0.250000000000$ $E$ $+00$  & 	\\
 1	& $+0.480675000000$ $E$ $+01$  &	$+0.480675000000$ $E$ $+01$  &	\\
 2	& $+0.424661478750$ $E$ $+02$  &       $+0.424661478750$ $E$ $+02$   &	\\
 3	& $+0.243939243681$ $E$ $+03$  & 	$+0.243939243680$ $E$ $+03$   &	\\
 4	& $+0.104714406827$ $E$ $+04$  &	$+0.104714276704$ $E$ $+04$ 	& $+0.00130123$\\
 5	& $+0.366000012391$ $E$ $+04$  &	$+0.365997458485$ $E$ $+04$   	& $+0.02553906$ \\
  6	&$+0.110533363334$ $E$ $+05$  & 	$+0.110532277550$ $E$ $+05$ 	&  $+0.1085784$ \\
 7	& $+0.300260377288$ $E$ $+05$  & 	$+0.300256729522$ $E$ $+05$ 	&  $+0.3647766$ \\
  8	&$+0.752398706312$ $E$ $+05$   &	  $+0.752389652239$ $E$ $+05$     &  $+0.9054073$  \\ [1ex] 
\hline 
\end{tabular}
\label{table:free_energy_all} 
\end{table}
\end{widetext}

\clearpage

\begin{table}
\caption{Uniform susceptibility series  per spin
with  all four $J_s$ (in Kelvin); 
$J_\pm = 0.754$, $J_{\pm\pm} = 0.487$, $J_{zz}= -0.290$ and $J_{z\pm}=-0.102$.
The $g$-tensor values are $g_\perp=6.05$ and $g_{zz}=2.5$.
The infinitesimal applied field is along the $[111]$ cubic direction.
The label $E$ $\pm m$ means $10^{\pm m}$. 
To convert into a susceptibility in $\text{ emu/mol}\cdot \text{K}$ units, 
the given values have to be multiplied by
$N_{\rm A} {\mu_{\rm B}}^2/k_{\rm B} \approx 0.375$
where  $N_{\rm A}$, ${\mu_{\rm B}}$ and $k_{\rm B}$ are the
Avogadro number, Bohr magneton and Boltzmann constant, respectively, all expressed in CGS units.
}
\centering 
\begin{tabular}{c c } 
\hline\hline 
 Order $n$ & expansion coefficient $a_n$ \\
 [0.5ex] 
\hline 
 0	&	$+0.662125000$ $E$ $+01$  \\
 1	&	$-0.988978150$ $E$ $+01$  \\
 2	&	$+0.683472045$ $E$ $+01$  \\
 3	&	 $-0.163940184$ $E$ $+01$ \\
 4	&	 $+0.131075702$ $E$ $+01$  \\
 5	& 	$-0.379184574$ $E$ $+01$   \\
 6	&	 $+0.103613852$ $E$ $+01$  \\ [1ex] 
\hline 
\end{tabular}
\label{table:free_energy_all} 
\end{table}


\begin{thebibliography}{30}
\bibitem{but_VCFE}
We omit here the tenuous energetic mechanism selection of $\psi_2$ 
discussed in Ref.~[\onlinecite{McClarty_ETO}]
and which involves a high-order van Vleck -like susceptibility.
\bibitem{dipoles_in_nn}
The  magnetostatic dipole-dipole interactions, written as
${\cal H}_{\rm D} = \sum_{i>j} (\mu_0 \mu_{\rm B}^2/4\pi) 
g_{\alpha\gamma} g_{\beta\delta}  S_i^\gamma S_j^\delta
 ( r_{ij}^{-3} \delta_{\alpha\beta} - 3r_{ij}^{-5} r_{ij}^\alpha r_{ij}^\beta)$,
have their nearest-neighbor part implicitly incorporated in the
$J_s$ parameters of ${\cal H}$ in Eq. \eqref{Hami}
and their contribution beyond nearest neighbors neglected.
\bibitem{Js_values}
The numerical values of those $J_s$ couplings are, 
(in Kelvin): $J_\pm=0.754\pm 0.087$, $J_{\pm\pm}=0.487\pm 0.058$, 
$J_{zz}=-0.290\pm 0.21$ and $J_{z\pm}=-0.102\pm 0.18$. 
The two largest couplings are of the $XY$ type and the physics is thus very different from spin ice (see 
for example Ref.\cite{Ross}).
\bibitem{Hz_form}
The components of ${\bm J}$ entering in ${\cal H}_{\rm Z}$ 
projected in the ground doublet are expressed as $J_i^\alpha = g_{\alpha\beta} S_i^\beta$
where $\alpha,\beta$ are cartesian components and sum over repeated indices is implied.
$g_{\alpha\beta}$ are the elements of the $g$-tensor with principal axes along the local $x$, $y$ and $z$
directions with eigenvalues $g_{xx}=g_{yy} \equiv g_\perp =5.97$
and $g_{zz} = 2.45$ from Ref.~[\onlinecite{Savary}].
\bibitem{move_critical_dipole} The addition of pseudo-dipolar
or indirect Dzyaloshinksii-Moriya interactions, such as those considered in
Ref.~[\onlinecite{McClarty_ETO}],  renormalize the critical value of the
magnetostatic dipolar interaction at which thermal ObD to $\psi_2$ occurs in
Ref.~[\onlinecite{Stasiak}] despite the ground state being Palmer-Chalker.~\cite{Palmer_Chalker}
\end{thebibliography}
\end{document}